\documentclass{aa}
\usepackage{amssymb}
\usepackage{times}
\usepackage{graphics}
\usepackage{psfig}

\def\asec{\ifmmode ^{\prime\prime}\else$^{\prime\prime}$\fi}
\def\lea{\ifmmode ^{<}_{\sim} \else $^{^{<}_{\sim}}$\fi}
\def\gea{\ifmmode ^{>}_{\sim} \else $^{^{>}_{\sim}}$\fi}
 
\begin{document}
\thesaurus{03.                  % A&A Section 3: Extragalactic astronomy
		(13.07.01;      % Gamma rays: bursts
		 11.16.1;)       % Galaxies: photometry
		}

\title{\bf $BVR_cI_c$  light curves  of GRB970508 optical remnant and
magnitudes of underlying host galaxy.}
\author{S.V. Zharikov,
V.V. Sokolov}
\offprints{S. Zharikov:zhar@sao.ru }

\institute{Special Astrophysical Observatory of R..A.S,
 Karachai-Cherkessia, Nizhnij Arkhyz, 357147 Russia; zhar,sokolov@sao.ru
}
 \date{Received / Accepted }

\authorrunning{Zharikov S.V. et al.}
 \titlerunning{The light curves and broand band spectrum of GRB970508 optical remnant.}

\maketitle

\begin{abstract}

The observations with the SAO-RAS 6-m telescope in July-August 1998 
show that the GRB970508 optical remnant has varied very little since November 1997. 
We can conclude that we observe a proper host galaxy 
without the GRB optical remnant. 
The fitting results of the light curve of the GRB remnant 
plus the host galaxy until 
about 400 days after the primary event are reported here. 
The best $\chi^2$-fits with a power law, for most of the filters used, are not 
acceptable. Some intrinsic variability of the GRB optical remnant fading is 
possible, which demands a more complex law presenting 
the largest multiband homogeneous data set from a single telescope.  

\end{abstract}

The GRB970508 OT is a second optical source related to a gamma-ray burst
 registered  by  the BeppoSAX satellite.
The optical variable object was first reported by H. Bond as a possible
optical counterpart of GRB970508 (Bond, 1997) and was independently
found in our data ($R_c$ band, 1-m telescope) only about 0.5 day later.
Observations of the GRB970508 optical remnant were continued
with the 6-m telescope in  the standard $BVR_cI_c$ bands
till Aug. 1998 (Sokolov et al., 1999).
\begin{figure*}
   \vbox{\psfig{figure=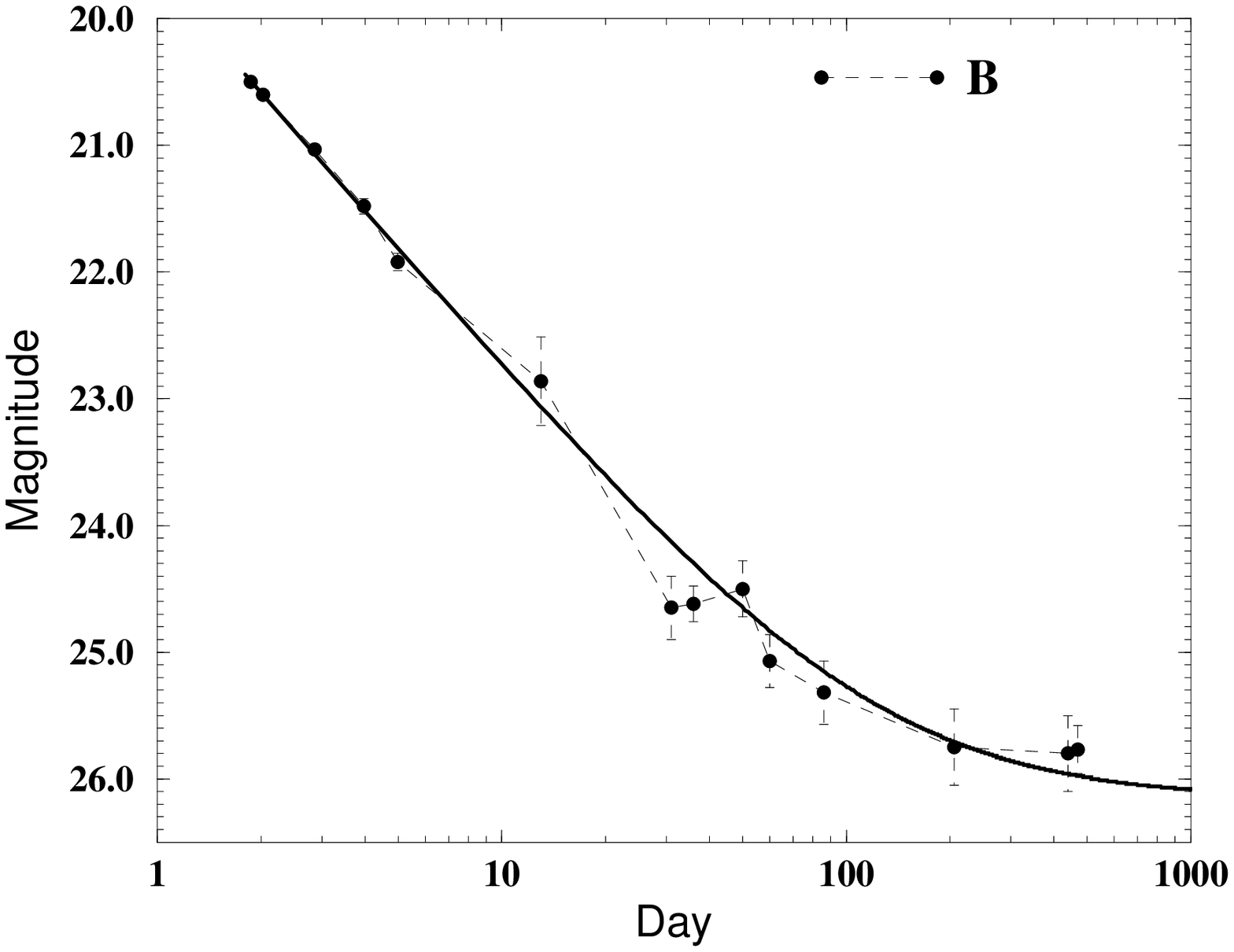,width= 7.5cm,%
 bbllx=40pt,bblly=40pt,bburx=580pt,bbury=450pt,clip=}}
   \vbox{\psfig{figure=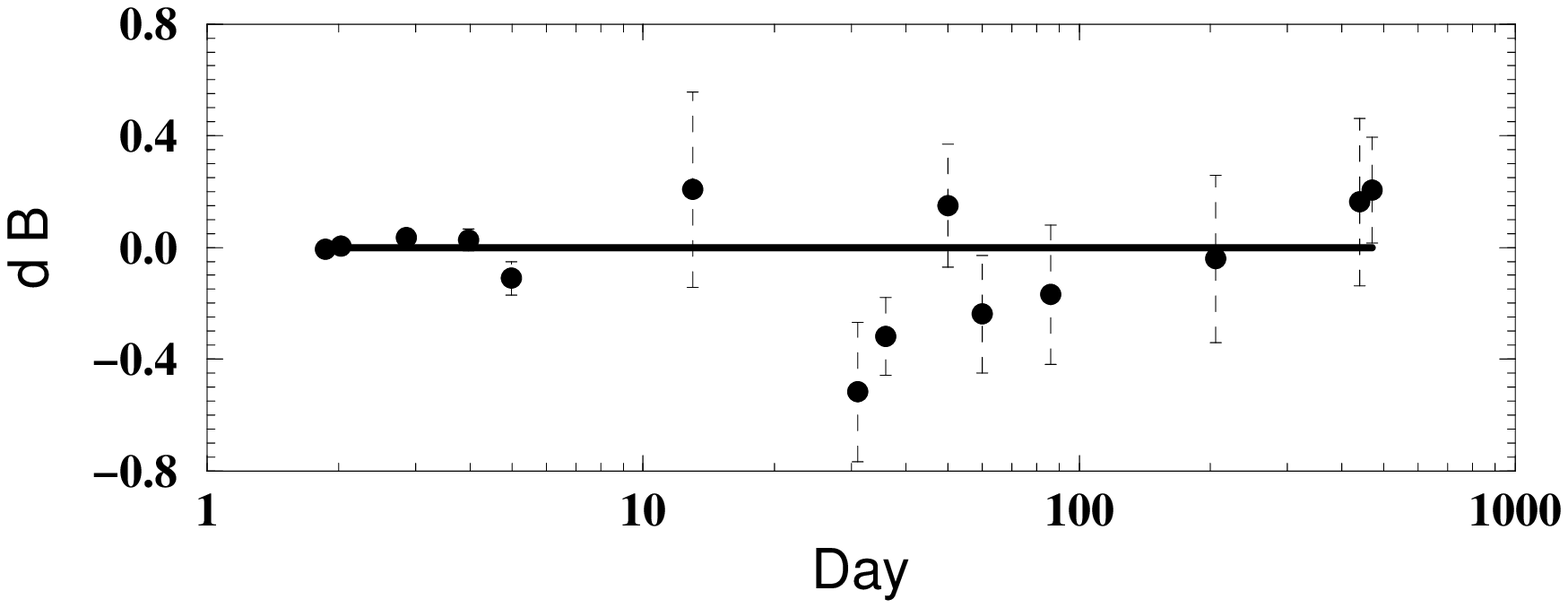,width=7.5cm,%
 bbllx=40pt,bblly=40pt,bburx=580pt,bbury=280pt,clip=}}
\vspace{-9.1cm}\hspace{8cm}
\vbox{
   \vbox{\psfig{figure=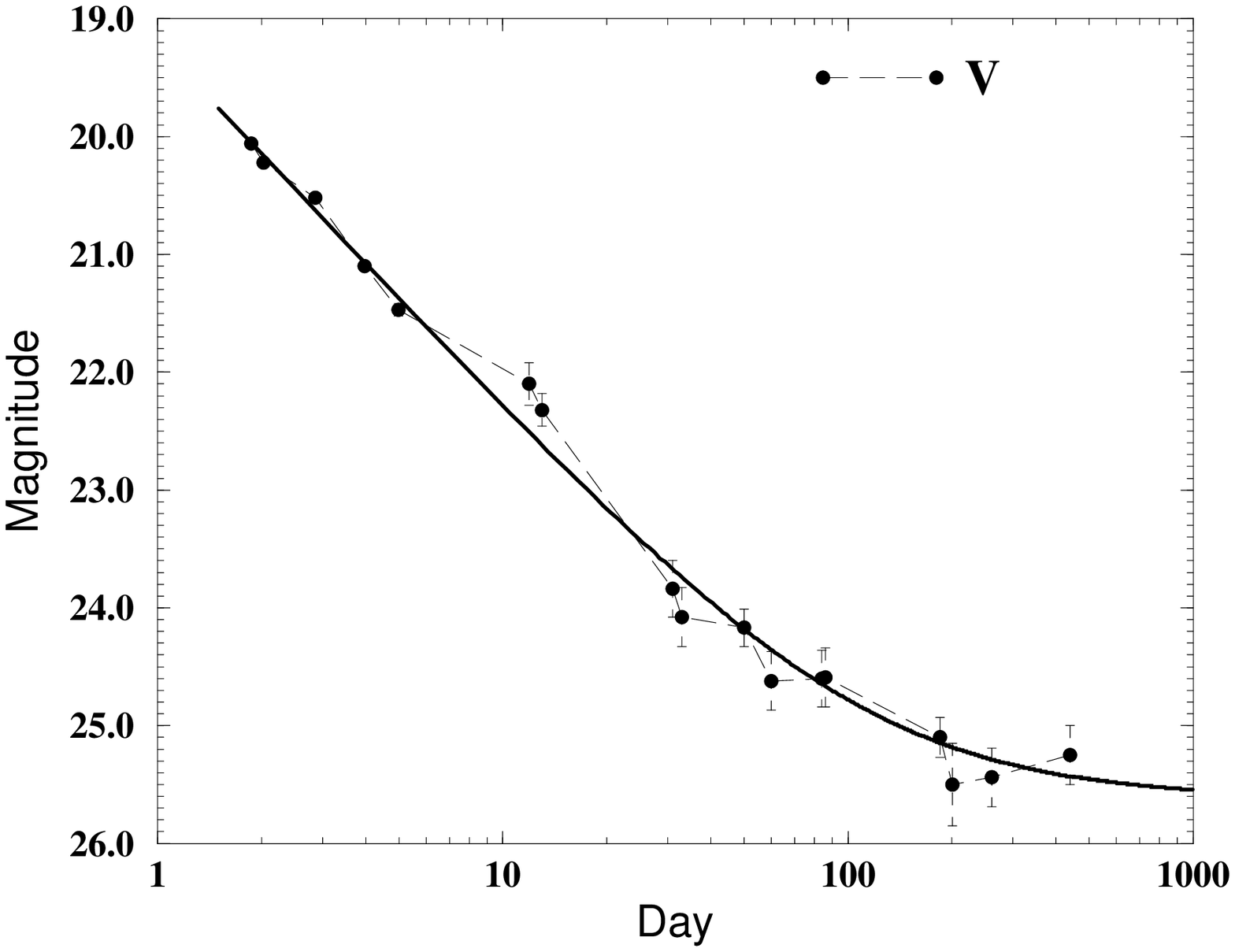,width= 7.5cm,%
 bbllx=40pt,bblly=40pt,bburx=580pt,bbury=450pt,clip=}}
   \vbox{\psfig{figure=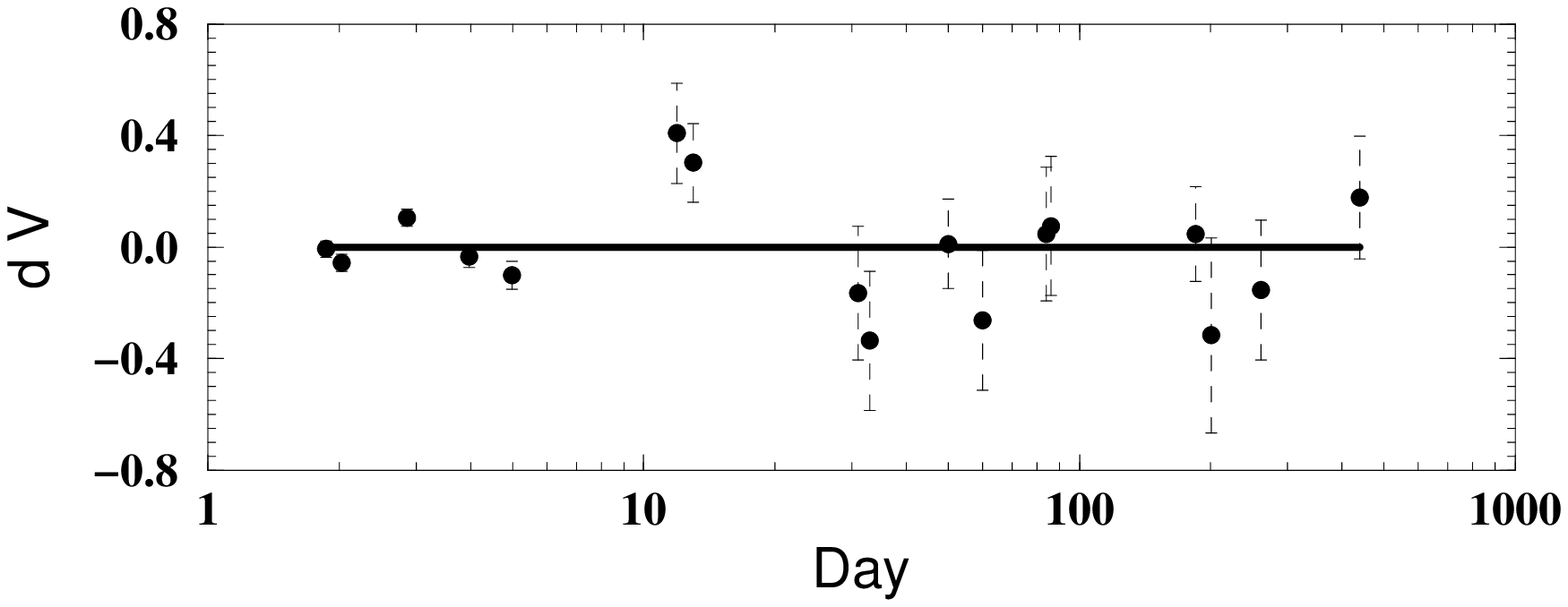,width=7.5cm,%
 bbllx=40pt,bblly=40pt,bburx=580pt,bbury=280pt,clip=}}
}
\vbox{
   \vbox{\psfig{figure=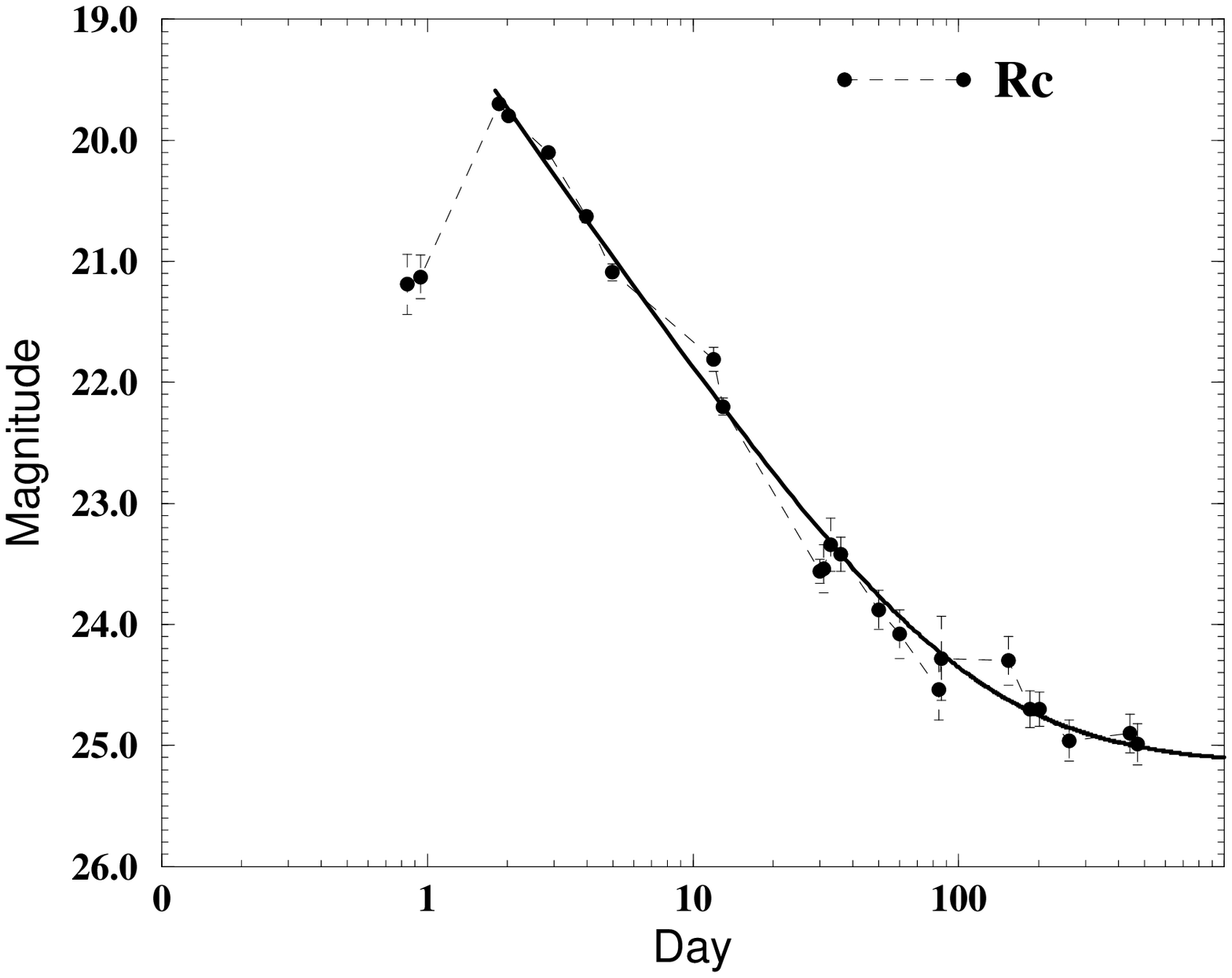,width= 7.5cm,%
 bbllx=40pt,bblly=40pt,bburx=580pt,bbury=450pt,clip=}}
   \vbox{\psfig{figure=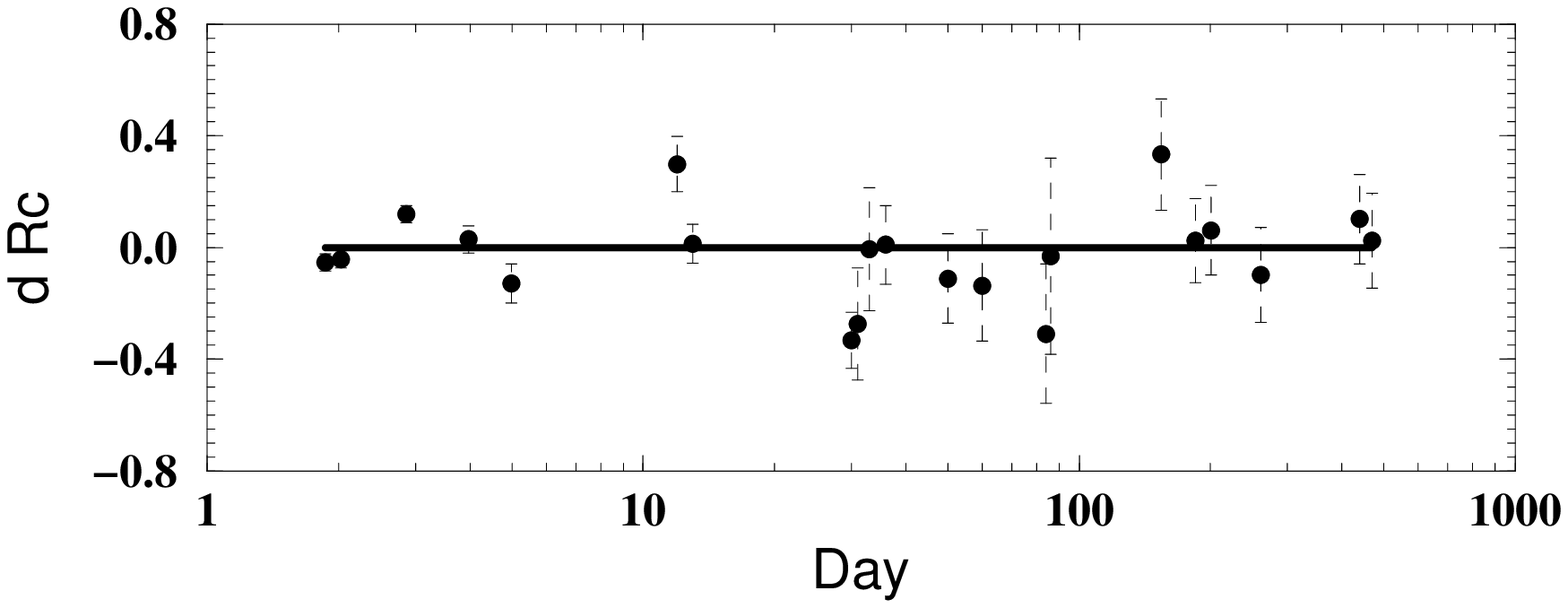,width=7.5cm,%
 bbllx=40pt,bblly=40pt,bburx=580pt,bbury=280pt,clip=}}
\vspace{-9.1cm}\hspace{10.5cm}
\vbox{
   \vbox{\psfig{figure=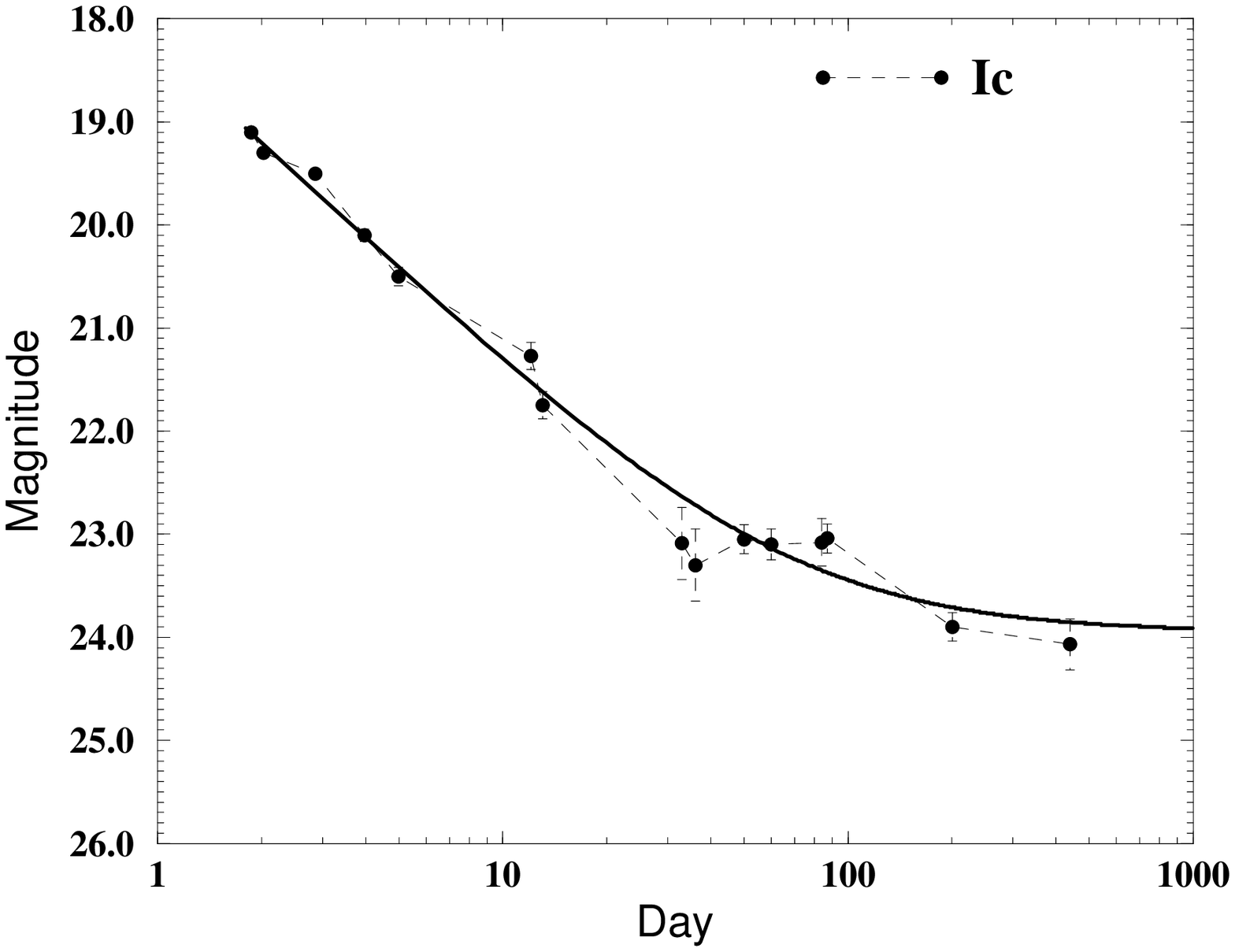,width= 7.5cm,%
 bbllx=40pt,bblly=40pt,bburx=580pt,bbury=450pt,clip=}}
   \vbox{\psfig{figure=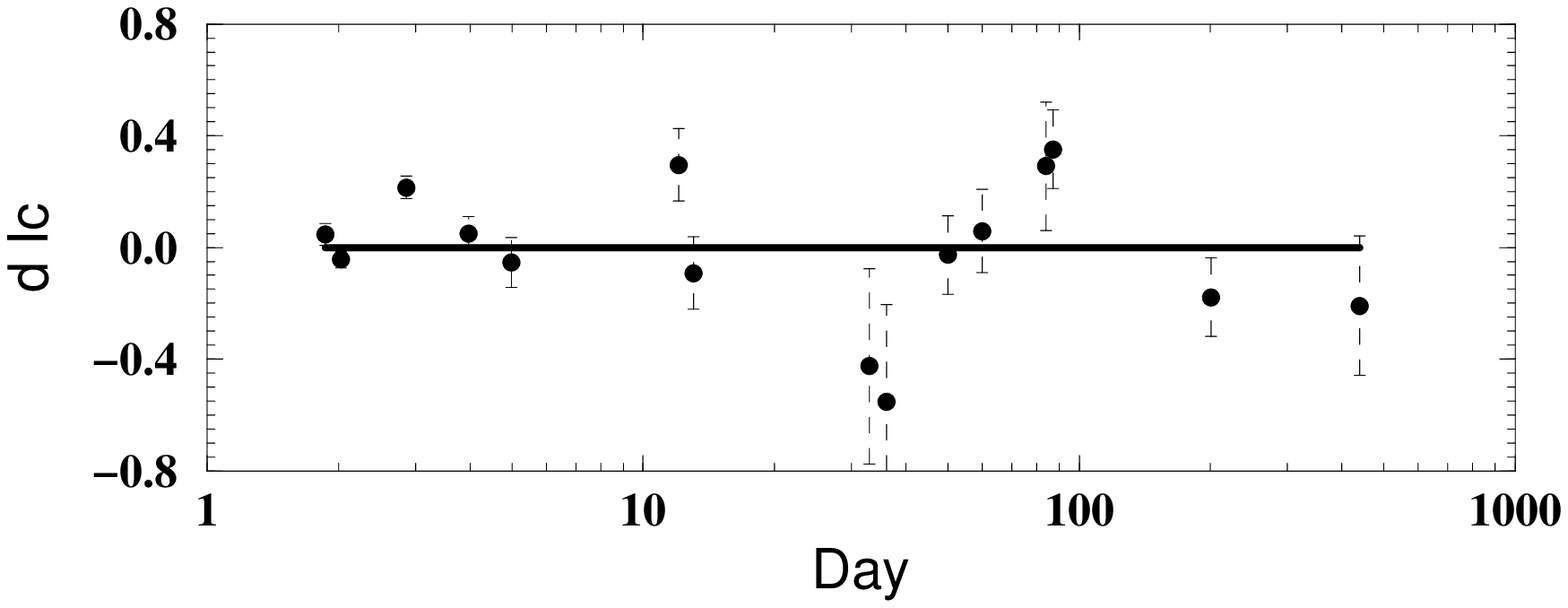,width=7.5cm,%
 bbllx=40pt,bblly=40pt,bburx=580pt,bbury=280pt,clip=}}
}
}
\caption{The $BVR_cI_c$ light curves of the GRB970508 optical remnant. 
The lines correspond to $F = F_o\times t^{\alpha}+F_C$ 
fits for each band, $t$ is time from the GRB trigger. 
The lower panels: the fits to the power law  plus
constants subtracted from the observational data. }
\end{figure*}

\begin{table}
 \caption[]{Magnitudes and fluxes for the host galaxy of GRB 970508.
{\bf (1)} The late-time observational magnitudes in Jul.-Aug. 1998. 
{\bf (2)} $\chi^2$ fits: $F = F_o\times t^{\alpha}+F_c$ with
$\alpha_B=-1.32 \pm 0.05$,
$\alpha_V=-1.24 \pm 0.07$,
$\alpha_{R_c}=-1.25 \pm 0.04$,
$\alpha_{I_c}=-1.18 \pm 0.07$.
{\bf (3)} The $\chi^2$ fits for $<\alpha>=-1.25 \pm 0.05$.
}
 \begin{tabular}{lllllll}
\hline
 Band & \multicolumn{1}{c}{Magnitude} & \multicolumn{1}{c}{log F$_c$}                                         & $\underline{\chi^2}$      \\
      &                               & \multicolumn{1}{c}{$\left(\frac{\rm erg}{{\rm cm}^2\ {\rm s}\ {\rm \AA}}\right)$} & (d.o.f)              \\
 & & & & \\ \hline
{\bf (1)} &&&& \\
    $B$ & 25.77 $\pm$ 0.19   &  $-18.52 \pm 0.07$&                 \\
    $V$ & 25.25 $\pm$ 0.22   &  $-18.54 \pm 0.09$&                 \\
    $R_c$ & 24.99 $\pm$ 0.17 &  $-18.66 \pm 0.07$&                 \\
    $I_c$ & 24.07 $\pm$ 0.25 &  $-18.58 \pm 0.10$&                 \\ \hline
{\bf (2)} &&&& \\
    $B$   & 25.99 $\pm$ 0.11 &  $-18.60 \pm 0.05$& 14.2/11         \\
    $V$   & 25.65 $\pm$ 0.17 &  $-18.70 \pm 0.07$& 36.6/14         \\
    $R_c$ & 25.16 $\pm$ 0.09 &  $-18.73 \pm 0.04$& 52.7/19         \\
    $I_c$ & 24.17 $\pm$ 0.28 &  $-18.62 \pm 0.11$& 44.3/12         \\ \hline
{\bf (3)} &&&& \\
    $B$ & 26.15 $\pm$ 0.16   &  $-18.67 \pm 0.06$&18.9/12          \\
    $V$ & 25.61 $\pm$ 0.16   &  $-18.69 \pm 0.07$&36.1/15          \\
    $R_c$ & 25.16 $\pm$ 0.09 &  $-18.73 \pm 0.04$&52.7/20          \\
    $I_c$ & 23.99 $\pm$ 0.25 &  $-18.55 \pm 0.10$&47.1/13          \\ \hline
\label{tab2}
 \end{tabular}
\vspace{-1cm}
 \end{table}

 The $BVR_cI_c$ magnitudes of the optical counterpart did not already change
  in magnitude error range during the last half-year  
 from November 1997 to August 1998. 
 We can conclude that we observe a pure
 host galaxy without the optical remnant of GRB970508.
 If the brightness changes within the errors,
 and the brighness decay of the optical remnant can still be described by
 a power-law relation for about a year after the burst occurence, 
 then the flux can be determined by fitting
 the observed $BVR_cI_c$ light curves with a two-component model,
 the sum of the GRB970508 optical remnant 
 fading according to a power-law, and a constant brightness
 of the host galaxy:
 $ F = F_o\times t^{\alpha}+F_c$.
 To investigate the possible variability of a faint source, one should 
 avoid any systematic shifts in the observational data
 due to different photometric systems in various instruments.
 That is why, for these fits 
 we used the homogeneous data set from the 6-m telescope only.
 In Table \ref{tab2} the host galaxy magnitudes and fluxes 
 are presented for 3 cases:
 (1) the proper late-time $BVR_cI_c$ observations in Jul.- Aug. 1998 
 without any fits, 
 (2) the fits with different slopes and corresponding $\chi^2/(d.o.f)$, 
 (3) the same $\chi^2$ fits with an average power-law slope 
  $<\alpha>=-1.25\pm0.05$.
  So, in cases (2) and (3) we indicated not observed, 
  but some theoretical values 
  for some model fits 
  and fluxes corresponding to $t$ tending to infinity.

The light curves for case (2) are presented in Fig.1.  
From large $\chi^2$ values we conclude that a single power law plus
a constant is not a good approximation of all observational data. 
On the other hand, we have shown (\cite{Sokolov}) that in the first 
days after the maximum 
the $\chi^2$ best fit was not a power law ($\chi^2 = 4.5/7$) 
but an exponential one ($\chi^2 = 0.97/7$). 
The identical exponential brightness fading   
was observed in 4 bands simultaneously. 
It is  
during the exponential flux decrease about 4 days after the maximum 
that the GRB970508 optical remnant has a stable power law spectrum with 
a slope $-1.10\pm0.08$ (\cite{M2},\cite{Sokolov}, \cite{Zharikov}).
The disregarding  of these 5 points in Fig.1 in all 
$BVR_cI_c$ bands or (which is the same) an arbitrary increase
of observational errors 3-4 times does not improve 
considerably the fitting with  
$ F = F_o\times t^{\alpha}+F_c$.
In particular, elimination of 5 points 
(with the smallest uncertainties, 0.03 - 0.07 mag)  
immediately after the brightness maximum in $R_c$ band in Fig.1  
yields $\chi^2/(d.o.f) = 16.8/14$.   
So a further intrinsic variability fading of the GRB 970508 
optical remnant is possible, 
which demands a more complex law for the presenting of all 
multiband $BVR_cI_c$ data set.
\acknowledgements{
The work was carried out under support of the
"Astronomy" Foundation (grant 97/1.2.6.4), INTAS N96-0315 and RBFI N98-02-16542.
}
 

\begin{thebibliography}{99}
\bibitem[Bond, 1997] {Bond}Bond H.E., 1997,  IAU Circ No.{\bf 6654}
\bibitem[Metzger et al., 1997]{M2}Metzger M. R., Djorgovski S.G., Kulkarni S. R., et al., 1997,  Nature, {\bf 387}, 879
\bibitem[Sokolov et al.,1998]{Sokolov}Sokolov V.V., Kopylov A.I., Zharikov S.V., Feroci M., Nicastro L., Palazzi E., 1998, A\&A., {\bf 334}, 117
\bibitem[Sokolov et al.,1999]{Sokolov1}Sokolov V.V., Zharikov S.V., Baryshev Yu.V. et al.,  1999, A\&A., {\bf 344}, 43
\bibitem[Zharikov et al.,1998]{Zharikov}Zharikov S.V., Sokolov V.V., and Baryshev Yu.V., 1998, A\&A., {\bf 337}, 356
\end{thebibliography}
 \end{document}